\documentclass[prd,aps,floats,twocolumn]{revtex4}
\usepackage{latexsym}
\usepackage{amssymb}
\usepackage[dvips]{graphicx}
\begin{document}
\newcommand{\beq}{\begin{equation}}
\newcommand{\eeq}{\end{equation}}
\newcommand{\ove}{\overline}
\newcommand{\half}{\frac 1 2 }
%
\newcommand{\et}{{\em et al}}
%
%
\newcommand{\Prd}{Phys.  Rev. D$\;$}
\newcommand{\Prl}{Phys.  Rev.  Lett.}
\newcommand{\Plb}{Phys.  Lett.  B}
\newcommand{\Cqg}{Class.  Quantum Grav.}
\newcommand{\Np}{Nuc.  Phys.}
\newcommand{\Grg}{Gen.  Rel.  \& Grav.}
\newcommand{\Fp}{Fortschr.  Phys.}
\newcommand{\Sch}{Schwarszchild$\:$}
\renewcommand{\baselinestretch}{1.2}

\title{Non-linear electrodynamics and the acceleration of the universe}

\author{M. Novello, S. E.  Perez Bergliaffa, and J. Salim}
\affiliation{Centro Brasileiro de
Pesquisas Fisicas, Rua Xavier Sigaud, 150, CEP 22290-180, Rio de
Janeiro, Brazil}
\vspace{.5cm}

\begin{abstract}
It is shown that the addition of a non-linear term to the Lagrangian of the electromagnetic
field yields a fluid with an asymptotically
super-negative equation of state, causing an accelerated
expansion of the universe. Some general properties of nonlinear electromagnetism in
cosmology are also discussed.
\end{abstract}
\date{\today}

\vskip2pc
\maketitle

\section{Introduction}

Observations of the redshift of supernovae type Ia \cite{riess}
and the cosmic microwave
background (CMB)\cite{cmb1} show that the universe is
undergoing a phase of accelerated expansion. The various
possibilities that have been examined to account for this
acceleration can be grouped in two classes. In the
models in the first class, matter with unusual properties is added to the
right-hand side of Einstein equations. Specifically,
this matter (dubbed \emph{dark energy}, although a  more appropriate name
may be \emph{smooth tension} \cite{carrolwhy})
must have a large negative pressure and, differently to the case of
dark matter, it must not cluster
with visible matter.
Amongst the most popular candidates for dark energy, we can
mention two. The first one is the cosmological constant
$\Lambda$, which yields accelerated
expansion due to the fact that $p_\Lambda = -\rho_\Lambda$. However, it is
strongly disfavoured
by the (120 orders of magnitude) disagreement between the observed value
and that naively predicted from quantum field theory \cite{cc, alam},
which implies a rather unnatural fine tunning.
The second candidate is the so-called quintessence
\cite{quint},
the simplest model of which is composed by a scalar field
\cite{kis}
coupled
to matter and under the influence
of an \emph{ad-hoc}
self-interaction potential \cite{othermodels}. The acceleration occurs in
a regime in which the
potential energy takes over the kinetic energy.
Although quintessence has been successful in
describing accelerated expansion, some problems remain:
the quintessence field has not been identified, and a derivation of the
potential from first principles is still lacking.

The second class of models that try to account for accelerated expansion
involve
a modification of the dynamics of the gravitational field in the
low-curvature regime. This modification can be achieved by
considering (in 5-dimensional scenarios)
the effects of the bulk on the dynamics of gravitation on the brane
\cite{dvali}, or by
directly adding to
the 4-dimensional gravitational action terms with negative powers of the
curvature scalar. An example of the latter
type is
given by the action \cite{carroll}
$$
S =\frac{M_{\rm Pl}^2}{2} \int \sqrt{-g} \left( R -
\frac{\alpha ^4}{ R} \right) d^4x,
$$
where $\alpha$ is a new fundamental parameter with units of mass. In this
article a third class of models will be introduced, which is a hybrid between the two
classes already mentioned. We shall assume that the action
for the
electromagnetic field
is that of Maxwell
with an extra term, namely
\beq
S = \int \sqrt{-g} \left( - \frac F 4+
\frac \gamma F \right) d^4x,
\label{action}
\eeq
where $F\equiv F_{\mu\nu}F^{\mu\nu}$. From the conceptual point of view,
this phenomenological
action has the advantage that it involves only the electromagnetic field,
and does not invoke entities that have not been observed (like
scalar fields) and/or speculative ideas (like higher-dimensions and
brane worlds).
At high values of the field invariant $F$,
the dynamics will be that of Maxwell plus corrections which are regulated by the
parameter $\gamma$, while at low values of $F$
it is the $1/F$ term that dominates \cite{gamma}.
As shown in the Appendix, this
modification of electromagnetism seems to be consistent with observation for small enough
values of $\gamma$. Notice also that the Lagrangian in
$S$ is gauge invariant. Hence charge conservation
is guaranteed in this theory.

The action given in Eqn.(\ref{action})
is actually only an example of a more general class
of actions for the electromagnetic field with Lagrangians that can be written as
$$
{\cal L} = \sum_k c_k F^k,
$$
where the sum may involve both positive and negative powers of $F$. We shall see that
in certain cases the ''electromagnetic fluid'' described
by this Lagrangian
can be thought of as composed by several noninteracting fluids, each
with an equation of state (EOS) of the type $p_i = w_i \rho_i$,
allowing for negative values of some of the $w_i$.

Taking the electromagnetic (EM)
field described by the action given in Eqn.(\ref{action}) as
source of Einstein's equations,
a toy model
for the evolution of the universe will be studied.
It will be shown that this model
displays accelerated expansion
caused by the dominance of the non-linear
EM term over other forms of matter. Let us recall that
under the parametrization $p = w\rho$ for the dark energy,
the universe will accelerate if $w<-1/3$. Values of $w$ beyond $w=-1$, which corresponds
to the cosmological constant, yield a super-negative EOS which violates
the dominant energy condition. However, observation allows for $w<-1$ \cite{mel},
so it is worthwhile considering models
with this type of EOS parameter.
Models with $w<-1$ in the literature are usually constructed
with
a scalar field with negative kinetic energy term \cite{caldwell1},
the so-called
phantom field. The evolution of a universe dominated by this type of matter
is very peculiar, ending in a \emph{Big Rip}
\cite{kam}: the energy density may grow without limit with time,
and there is a singularity in the future. As we shall see, the
non-linear electromagnetic model studied here
yields ordinary radiation plus a dark energy component
with $w<-1$.

\section{Nonlinear electromagnetism in Cosmology}

The effects of a nonlinear electromagnetic theory in a cosmological setting have
been studied in several articles. The authors of \cite{vitorio} showed that nonlinear
corrections coming from an Euler-Heisenberg-like Lagrangian
can be important in the very early universe, leading to the
avoidance of the singularity. Some attention has been devoted to the dynamics of
a universe governed by matter described by
Born-Infeld
theory \cite{portuga1}. In particular,
it has been shown that
under certain assumptions (as for instance the existence of a compactified space)
a Born-Infeld field as a source
yields accelerated expansion \cite{portuga1}.

We shall study first some general properties of nonlinear electrodynamics
in cosmology, and then pay attention to the specific case
of the action given in Eqn.(\ref{action}). Let us
start with the theory defined by the Lagrangian
\beq
{\cal L} = \sum_k c_k F^k,
\label{action2}
\eeq
where the sum can have both positive and negative powers of the field invariant.
Due to the isotropy of the spatial
sections of the Friedman-Robertson-Walker (FRW)
model, an average procedure is needed if electromagnetic fields are to
act as
a source of gravity \cite{robertson}. Let us define first the volumetric spatial
average of a quantity $X$ at the time $t$ by
\beq
\overline X \equiv \lim_{V\rightarrow V_0} \frac 1 V \int X \sqrt{-g}\;d^3x,
\eeq
where $V = \int \sqrt{-g}\;d^3x$ and $V_0$ is a sufficiently large time-dependent three-volume.
In this notation, the electromagnetic field can act as a source for the FRW model
if \footnote{We are using the sign convention $(+---)$ for the metric.}
\beq
\overline E_i =0, \;\;\; \overline B_i =0,\;\;\; \overline{E_i B_j}=0,
\eeq
\beq
\overline{E_iE_j}=-\frac 1 3 E^2 g_{ij}, \;\;\;\overline{B_iB_j} = -\frac 1 3 B^2 g_{ij}.
\eeq
With these conditions, the energy-momentum tensor of the EM field associated to
Maxwell's Lagrangian can be written as
\beq
\ove{T}_{\mu\nu} = (\rho + p) v_\mu v_\nu - p\; g_{\mu\nu},
\label{tmunu}
\eeq
where
$$
\rho = 3 p =\half (E^2 + B^2).
$$
Under the same assumptions, a general nonlinear Lagrangian yields the
stress-energy tensor
given by Eqn.(\ref{tmunu}) with
\beq
\rho = -L-4E^2L_F,\;\;\;\;p = L+\frac 4 3 (E^2-2B^2)L_F,
\label{emse}
\eeq
where $L_F\equiv\frac{dL}{dF}$.
As we shall see below, the case in which $E^2=0$ is the only relevant case in cosmology.
It is straightforward to prove from Eqn.(\ref{emse})
that when there is only a magnetic
field, the fluid can be thought of as composed of a collection of
non-interacting fluids indexed by $k$, each of which obeys the equation of state
\beq
p_k = \left( \frac{4k}{3} - 1 \right) \rho_k.
\label{partialp}
\eeq
In particular, the fluid coming from Eqn.(\ref{action}) is composed of ordinary radiation
with $p_{(1)} = \frac 1 3\; \rho_{(1)}$ and of another fluid with EOS
$$
p_{(-1)} = -\frac 7 3 \;\rho_{(-1)}.
$$
It is precisely this component with negative pressure that may drive the
acceleration of the universe.

Let us remark that since we are assuming that
$\overline{B}_i =0$, the magnetic field
induces no directional effects in the sky,
in accordance with the symmetries of the standard cosmological model.

From Einstein's equations, the acceleration of the universe is related
to its matter content by
\beq
3 \frac{\ddot a}{a} = - \half (\rho + 3 p).
\label{acc}
\eeq
In order to have an accelerated universe, the matter must satisfy the
constraint $(\rho + 3 p)<0$. Assuming that
the non-linear electromagnetic field is the dominant source of gravity
and using the quantities defined in Eqn.(\ref{emse})
we can write
$$
\rho + 3 p = 2(L-4B^2L_F).
$$
Hence the constraint $(\rho + 3 p)<0$
translates into
\beq
L_F > \frac{L}{4B^2}.
\eeq
It follows that any nonlinear electromagnetic theory that satisfies this inequality
yields accelerated expansion under the conditions given in Eqns.(4) and (5).

As an aside, note that from Eqn.(\ref{emse}),
\beq
\rho + p = -\frac 8 3 \;(E^2 + B^2) L_F.
\eeq
Consequently, nonlinear electromagnetic theories
that have a \emph{state} for which $L_F = 0$
may generate an effective cosmological constant.

\subsection{A particular example}

We shall consider next the action given by Eqn.(\ref{action}) in the cosmological
setting given by the FRW
metric. The corresponding equations of motion (which reduce to
ordinary electromagnetism when $\gamma = 0$)
are
\beq
\left[ \left( 1 + \frac{4\gamma}{F^2}\right) F^{\mu\nu}\right]_{;\nu}=0.
\label{eom}
\eeq
From now on we shall assume that $E^2 = 0$, while there is a residual
magnetic field characterized by
$\overline B_i =0, \overline{B_iB_j} = -\frac 1 3 B^2 g_{ij}$.
This assumption is consistent with the fact that
the electric field will be screened by the charged primordial plasma,
while the magnetic field lines will be frozen \cite{lemoine}.
In this case,
$$
\rho = \frac{B^2}{2} + \frac{\mu^8}{2} \frac{1}{ B^2},
$$
where we have set
$\gamma \equiv - \mu^8$ \cite{gamma}.
From the conservation law
$$
\dot \rho + 3(\rho + p ) \frac{\dot a}{a} = 0,
$$
and setting $a_0=1$
we get that $B = B_0 / a^2$. Hence the evolution of the density with the scale factor is
\beq
\rho = \frac{B_0^2}{2}\; \frac{1}{a^4} + \frac{\mu^8}{2B_0^2}\; a^4.
\label{density}
\eeq
For small $a$ it is the ordinary radiation term that dominates. The $1/F$ term takes over
only after $a=\sqrt{B_0}/\mu$, and grows without bound afterwards. In fact, the curvature
scalar is
$$
R = T^\mu_{\;\mu} = \rho-3p = \frac{4\mu^8}{B_0^2}\; a^4,
$$
showing that there is a curvature singularity
in the future of universes for which $a\rightarrow\infty$
(see Sect.\ref{dynamics}).

Using the expression for the density in Eqn.(\ref{acc}) gives
$$
3 \frac{\ddot a}{a} +  \frac{B_0^2}{2}\; \frac{1}{a^4} - \frac 3 2
\frac{{\mu^8}}{B_0^2}\; a^4=0.
$$
To get a regime of accelerated expansion, we must have
$$
 \frac{B_0^2}{a^4} -  3
\frac{{\mu^8}}{B_0^2}\; a^4 < 0,
$$
which implies that the universe will accelerate for $a>a_c$, with
$$
a_c = \left(\frac{B_0^4}{3\mu^8}\right)^{1/8}.
$$
We have shown that when the matter content of the universe
is dominated by a non-linear EM field with the
EOM (12)
the universe enters a phase of accelerated expansion.
The non-linear theory depends upon the yet undetermined
parameter $\mu$.
It is possible to get an order-of-magnitude estimation for this parameter as follows.
Recent observations \cite{spergel} suggest that the approximate current values for
$\Omega_{de}$ (the
ratio of the dark energy density to the critical density
$\rho_c$)
and the Hubble parameter are $\Omega_{de}\approx 0.70$
and $h  \approx 0.70$. Assuming that the dark energy can be described by
the
non-linear term in Eqn.(1), we get that
$\rho_{(-1)} \approx 0.70 \rho_c$. In other words,
\beq
\frac{\hbar^2\mu^8}{B_0^2} \approx 1.40 \rho_{c}.
\label{mu}
\eeq
From this equation, we can get an upper bound for $\hbar\mu^4$ if we take $B_0$
as that associated to the CMB radiation. This can be obtained
from
$\Omega_{rad} = 2.47\times10^{-5}\;h^{-2}$. It follows that
$B_0^2 = 2.47\times10^{-5}\;h^{-2} \rho_{c}$. Inserting this value in Eqn.(\ref{mu})
we get
\beq
\hbar\mu ^4\approx 3.74\times 10^{-28}\textrm{ gr/cm$^3$}.
\eeq
This value of
$\hbar\mu ^4$ together with
the results given in the appendix show that, in the case of a point charge,
the corrections to Maxwell's
electromagnetism coming from the non-linear term in Eqn.(\ref{action})
are negligible.
\subsection{Dynamics}
\label{dynamics}
The dynamics of the universe with matter density given by Eqn.(\ref{density})
can be obtained qualitatively from the analysis of Einstein's equations
We shall be interested only in the early regime (where radiation with
$p = 1/3\rho$ drives the expansion), and in the late evolution (where the 1/F term is
important). Consequently, dust-like matter will be disregarded.
Equation
$$
\frac{\ddot a}{a}=-  \frac{B_0^2}{6}\; \frac{1}{a^4} + \frac 1 2
\frac{{\mu^8}}{B_0^2}\; a^4,
$$
implies that $\ddot a$ has a zero at $a=a_c$, while equation
$$
\frac k 3 \rho = \left(\frac{\dot a}{a}\right)^2 + \frac{\epsilon}{a^2}
$$
where $k=8\pi G$, and $\epsilon=0,\pm 1$ depending on the 3-section of the geometry,
can be written as
\beq
\dot a ^2 + V(a) = -\epsilon,
\label{part}
\eeq
where
$$
V(a) = -\frac k 6 \left( \frac{B_0^2}{a^2} + \frac{\mu^8}{B_0^2}a^6\right).
$$
Eqn.(\ref{part}) can be interpreted as describing the one-dimensional
motion of a particle with energy $-\epsilon$ under the influence of the potential $V(a)$.
The potential is always negative, and has a maximum at $a=a_c$.

The analysis of the flat and open cases shows that this system allows for
universes that starting with a big bang at $a=0$, expand with negative acceleration
up to $a=a_c$, and then run away to infinity with positive acceleration, the difference
between the two being the value of the Hubble parameter at $a=a_c$. Note in particular
that for $\epsilon = 0$, the dynamics is invariant under the change $a\rightarrow 1/a$.

\begin{figure}
\begin{center}

\includegraphics[width=0.55\textwidth]{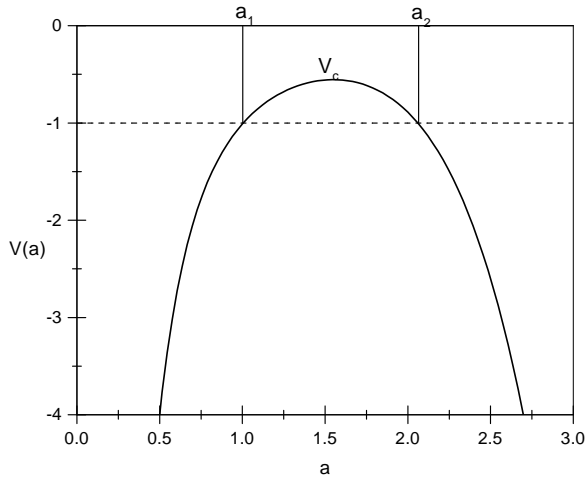}
\caption{Qualitative plot of the potential for the case $V_c > -1$.
The region between $a_1$ and $a_2$ is classically forbidden.}
\label{effpot1}
\end{center}
\end{figure}

The closed case
is more interesting due to the fact that
\beq
V(a_c)\equiv V_c = -\frac{2}{3}^{7/4}k B_0 \mu^2
\eeq
can be $\lessgtr -1$ \cite{ellis}. We shall examine three separate cases. \\[.3cm]
\emph{Case (1):} $V_c < -1$. This case displays the same behaviour as that
of the $\epsilon =1,-1$ cases: the universe may start from a big bang, then a deaccelerated
expansion follows up to $a=a_c$, and afterwards accelerated eternal expansion takes place.
\\[.2cm]
\emph{Case (2):} $V_c = -1$. The same as the previous case, but this time
$H(a_c) =0$, so that $a=a_c$ is an unstable equilibrium point.\\[.2cm]
\emph{Case (3):} $V_c > -1$. Two different behaviours are allowed (see Fig.(\ref{effpot1})).
The universe
may start from $a=0$, expand up to $a_1$ and then re-collapse. Alternatively, it may start
from infinity, collapse to $a=a_2$, and then re-expand to infinite.

\section{Discussion}

The accelerated expansion suggests that the EOS parameter $w$
of the dark energy may be smaller that -1, the limit
usually imposed by theoretical arguments such as the dominant energy condition.
The confirmation of these findings will have a profound impact for fundamental physics.
In this article
we have given a particular realization of a model that displays a time-dependent EOS,
with a super-negative
EOS parameter as a limit.
Instead of using a negative kinetic energy term for a scalar field, the price to pay
was to modify the action for the electromagnetic field by the addition of a non-linear term
that respects gauge invariance.
As shown in the appendix, this modification
is harmless in the case of a point charge \footnote{The correction can be safely neglected
also
in the case of the energy spectrum of the hydrogen atom (E. Elbaz, private communication).}.
The result of the modification in FRW
cosmology is to accelerate the universe, ending in a Big Rip, independently of
the spatial section of the geometry (except for one particular case
in the closed universe, see Fig.(1)).

Attempts are currently under way to
reconstruct the EOS from observation \cite{alam}. This reconstruction
implies the measurement of the \emph{jerk} (third derivative of the scale factor)
\cite{visser}.
Although the toy model we analyzed here yields an EOS
with a value ($\omega_{(-1)} = -7/3$) that seems to be
rather disfavored by observation \cite{kam, alam},
it would be interesting to see if
the addition of other non-linear terms
gives the chance of
obtaining more
realistic values for the EOS parameter.

\section*{Appendix}

The least we should require of any modification to Maxwell's equations
is consistence with the case of a static electric field
generated by a point charged particle. The EOM for a general
non-linear Lagrangian
$L=L(F)$ are
\beq
\partial_\mu (L_F\;F^{\mu\nu})=J^\nu,
\label{nleom}
\eeq
where $J^\nu$ is an external current. It follows immediately that
$\partial_\nu J^\nu = 0$, showing again that charge is conserved
in a theory with $L=L(F)$. In the case of the point charge, Eqn.(\ref{nleom})
reduces to
$$
r^2 L_F\;E(r)= {\rm const.}.
$$
In the case of the lagrangian given in Eqn.(\ref{action}) we get
\beq
E(r)^4 + \frac{Q}{r^2} E(r)^3  - \mu^8 = 0,
\label{ef}
\eeq
where $Q = 4\times {\rm const.}$, and we have set $\hbar =1$.
Although this cubic equation could be solved
exactly, we shall only need its solution for large and small values of $r$.
For $r^2>>Q/\mu^2$ we get
\beq
E(r) = \mu^2 - \frac 1 4 \frac{Q}{r^2} + O\left(\left(\frac{Q}{r^2}\right)^2\right),
\label{limit1}
\eeq
while in the limit $r^2<<Q/\mu^2$,
\beq
E(r) = -\frac{Q}{r^2} \left( 1 - \left(\frac{\mu^2 r^2}{Q}\right)^4 \right)+ O\left(
\left(\frac{\mu^2 r^2}{Q}\right)^8\right).
\label{limit2}
\eeq
By taking derivatives of Eqn.(\ref{ef}) it can be shown that
the function $E(r)$ has no extrema \footnote{Note that $E=0$ is not
a solution of Eqn.(\ref{ef}).}. Hence, the modulus of the electric field monotonically
decreases with increasing $r$, from an infinite value at the origin
to a constant (nonzero but small) value at infinity. This situation is akin to that
in the theory defined by the action
$$
S =\frac{M_{\rm P}^2}{2} \int \sqrt{-g} \left( R -
\frac{\alpha^4}{ R} \right) d^4x.
$$
It was shown in \cite{carroll} that the static and spherically symmetric solution
of this theory does not approach Minkowski asymptotically; it tends instead to
(anti)-de Sitter spacetime. Analogously to the gravitational
field in \cite{carroll}, by choosing the
parameter $\mu$ small enough the electric field of a point particle in
the non-linear theory
given by Eqn.(\ref{action}) will pick up only extremely tiny corrections, given by
Eqns.(\ref{limit1}) and (\ref{limit2}).

\section*{Acknowledgements}

SEPB acknowledges FAPERJ for financial support. JS is supported by CNPq.
MN acknowledges support of FAPERJ and CNPq.


\begin{thebibliography}{99}
\bibitem{gamma} Notice that on dimensional grounds,
$
\gamma = \hbar ^2\times \mu^{8},
$
where $\mu$ is a fundamental constant with dimension of length$^{-1}$.
\end{thebibliography}
\end{document}